\documentstyle[graphics,aps,amsmath,epsf]{revtex} 

\begin{document}
%\draft
%%%%%%%%%%%%%%%%%%%%%%%%%%%%%% User specified LaTeX commands.
%%%%%%%%%%%%%%%%%%%%%%%%%%%%%%
%\begin{document}
\title{Center-of-Mass Properties of the Exciton in Quantum Wells}
\author{A. Siarkos, E. Runge, and R. Zimmermann}
\address{Institut f\"ur Physik, Humboldt-Universit\"at zu Berlin, Hausvogteiplatz 5-7, 10117 Berlin, FRG}
\date{\today}
\maketitle
\begin{abstract}
We present high-quality numerical calculations of the exciton center-of-mass
dispersion for GaAs/AlGaAs quantum wells of widths in the range \( 2-20 \)~nm.
The \( \vec{k}\cdot \vec{p}\,  \)-coupling of the heavy- and light-hole bands
is fully taken into account. An optimized center-of-mass transformation enhances
numerical convergence. We derive an easy-to-use semi-analytical expression for
the exciton groundstate mass from an ansatz for the exciton wavefunction at
finite momentum. It is checked against the numerical results and found to give
very good results. We also show multiband calculations of the exciton groundstate
dispersion using a finite-differences scheme in real space, which can be applied
to rather general heterostructures.
\end{abstract}
%%%%%%%%%%%%%%%%%%%%%%%%%%%%%%%%%%%%%%%%%%%%%%%%%%%%%%%%%%%%%%%%%%%%%%
%\begin{multicols}{2}
%\narrowtext
%%%%%%%%%%%%%%%%%%%%%%%%%%%%%%%%%%%%%%%%%%%%%%%%%%%%%%%%%%%%%%%%%%%%%%
\section{Introduction}

Excitons dominate the optical properties of low-dimensional semiconductor heterostructures
such as quantum wells (QW) and quantum wires. The relative motion of the constituent
particles and their center-of-mass (COM) motion determine different characteristics
of the optical spectra and exciton kinetics. 

The exciton relative motion in QW is well studied and understood. The confinement
of the carriers along one or two spatial directions into regions 
comparable to or smaller than the bulk exciton size enhances the
effect of the electron-hole Coulomb interaction. This results in larger binding
energies and oscillator strengths and in an increased stability compared to
bulk excitons. Therefore, excitons are observed even at room temperature in
these structures. The effect of the reduced dimensionality is as a rule much
larger on the exciton groundstate than on the excited states.

Details of the excitonic optical spectra of QW related to the COM~motion
like, e.g., inhomogeneous broadening and Stokes shift between photoluminescence
(PL) and absorption are frequently used for structure characterization. These
features are influenced by exciton localization and diffusion in the presence
of interface or alloy disorder \cite{Runge}. Optical spectra and their temporal
evolution are determined by the exciton formation processes \cite{exciton formationi}
and the subsequent energy and spin \cite{spin relaxation} relaxation dynamics.
Spatially resolved spectroscopy techniques like micro PL and near-field scanning
optical microscopy allow direct observation of exciton COM~quantization in
local potential minima \cite{Hess}. All these phenomena are intimately related
to the exciton COM~properties whereby different energy and COM~momentum
regions of the exciton dispersion are probed in different processes. 

In many heterostructure systems of interest like, e.g., GaAs/\-AlAs,
InGaAs/\-InP and ZnCdSe/\-ZnSeS, the exciton can be described in the effective-mass
approximation (Wannier exciton) due to its small bulk binding energy (e.g.,
4~meV for GaAs). In this approximation, the degeneracy of the valence bands at
the center of the Brillouin zone for materials of cubic or zinc-blende symmetry
was first taken into account by Dresselhaus \cite{Dressel}. 
He also pointed out the 
absence of a well-defined COM~transformation due to this degeneracy. 
Altarelli and Lipari \cite{Altar} calculated the exciton COM~dispersion
for direct- and indirect-gap bulk semiconductors. They demonstrated that the
ambiguity in the choice of the COM~transformation can be used to achieve
formal simplicity or optimal numerical convergence. For bulk GaAs, where the
heavy- to light-hole mass ratio is large, the exciton dispersions are found
to be strongly anharmonic and show avoided crossings between different branches. 

In semiconductor QW, the broken translational symmetry in the growth direction
leads to the splitting of heavy and light holes at the \( \Gamma  \)-point,
and subsequent formation of heavy- and light-hole excitons. 
Due to the large hole-to-electron
mass ratio, the influence of the valence-band mixing on the COM~motion is
greater than on the relative motion. The exciton dispersions are, thus, strongly
non-parabolic. Direct consequences of the exciton dispersion anharmonicity in
QW like slow indirect excitonic transitions due to camel-back shaped dispersions
\cite{52} have been experimentally observed \cite{53}.

The multiband exciton (i.e., with the full coupling of heavy- and light-hole
bands taken into account) has been theoretically and numerically thoroughly
investigated at vanishing COM~momentum \( Q \) 
\cite{BS85,Broido/Sham,Sander/Chang,Andreani/Pasq,Choa/Chuang,Zhu/Chang,Chen/Bajaj,Winkler}. 
The numerical effort for such calculations remains reasonable due to the high
symmetry of this point. In contrast, only a few publications on multiband calculations
of exciton COM~dispersions in QW exist (\cite{JorRoesBro,Triques,US}) 
since these are very demanding. Methods for improving the numerical
accuracy and reducing the effort of such calculations are clearly necessary.
Particularly useful would be an easy-to-use approach that gives the main features
of the exciton dispersion with at least moderate accuracy.

The main focus of the present work lies on the exciton groundstate dispersion
and its properties. A secondary goal is to study the feasibility of numerical methods
for calculating the exciton groundstate dispersion in more complicated structures
like V-groove quantum wires \cite{Kapon}. Excitons in GaAs/Al\( _{0.3} \)Ga\( _{0.7} \)As
symmetric QW are considered (section \ref{SMODEL}). Two different methods for
the dispersion calculation are used: (i) the extension for \( Q\neq 0 \) of
the well-known \cite{Choa/Chuang} expansion of the exciton in the product space
of electron and hole subband states (section \ref{SKSPACE}), and (ii) a finite-differences
scheme in real space (section \ref{REALSPACE}) with a groundstate-adapted Coulomb
discretization \cite{US}. Method (i) gives high-quality numerical results but
is not feasible, e.g., for quantum wires of complicated geometry. With method
(ii) the groundstate exciton dispersion in a V-groove quantum wire is tractable.
Its convergence properties are checked here against the results of (i). Results
for V-groove quantum wires will be presented elsewhere. 

Improving own previous results \cite{US}, we address in detail the problem
of the choice of the COM~transformation and introduce an optimized, groundstate
adapted COM~transformation that greatly enhances the numerical accuracy and
stability of our results (section \ref{CMM}). 

Quasi as a by-product, a semi-analytical expression for the average exciton
groundstate mass suitable for exciton localization problems is derived (section
\ref{SMASS}). This expression is of great practical importance since it gives
for not too wide QW reliable mass values. The only necessary ingredients are the
lowest subband dispersion and a good estimate for the groundstate Bohr radius. 

Finally, we discuss the results of the exciton dispersion calculations in momentum-
(section \ref{KSECTION}) and real-space (section \ref{RSECTION}). The results
of our semi-analytical expression for the average exciton groundstate mass are
compared to various other mass expressions as well as with the numerical
dispersions separately (section \ref{MASSSECTION}).

\section{Theoretical model\label{SMODEL}}

We consider the well-studied system of direct Wannier
excitons in a single symmetric GaAs/Al\( _{0.3} \)Ga\( _{0.7} \)As type I quantum
well grown in \( \langle 100\rangle  \) direction. Many aspects of the presented
results can be effortlessly extended to Wannier excitons in other, more general
heterostructures. 

In the envelope function approximation, the Wannier exciton is described by
the Hamilton operator 
\begin{equation}
\label{efmHam}
\begin{array}{ccl}
H & = & H_{e}(\vec{r}_{e})+H_{h}(\vec{r}_{h})+V_{Coul}(\vec{r}_{e}-\vec{r}_{h})\, \, .
\end{array}
\end{equation}
 \( H_{e,h} \) describe the material-dependent bandstructure of the respective
particles in the vicinity of the \( \Gamma  \)-point, and \( V_{Coul} \) stands
for the attractive Coulomb interaction. We choose the coordinate system as usual
with the \( z \)-axis in growth direction \( \langle 100\rangle  \); 
\( \vec{r}_{e}=\left( x_{e},y_{e},z_{e}\right) ,\, \vec{r}_{h}=\left( x_{h},y_{h},z_{h}\right)  \)
denote the space coordinates of electron and hole, respectively. For the materials
involved, the conduction band is to a good approximation parabolic; anharmonicities
in the conduction band arise mainly through the interaction with the light and
split-off valence bands which is small due to the relatively large band gaps.
The valence band is adequately described by the Luttinger Hamiltonian \cite{Lut}
in the axial approximation \cite{BS85,Zhu/Huang}, which takes into
account explicitly the coupling of the heavy- and light-hole bands (\( \Gamma _{8}^{v} \))
but suppresses warping. The coupling to the split-off (\( \Gamma _{7}^{v} \))
band can be safely neglected for subband states with energies up to approximately
50~meV from the band edge because of the relatively large energy separation.
We neglect the effect of the different dielectric constants (no image charge
effects) \cite{Andreani/Pasq,Chen/Bajaj,Thoai/Zim},  and also
all effects that lead to a small spin-splitting like lack of inversion symmetry
of the bulk material \cite{Zhu/Chang} or the interfaces \cite{spin splitting}
as well as the exchange part of the Coulomb interaction \cite{JorRoesBro}.
The electron spin is irrelevant and will be given a fixed value of \( +1/2 \)
in the present work. The quantization axis of the electron spin and of the hole
angular momentum \( J \) is taken along the growth direction, and we use for
the valence band edge states the same convention as in Ref.~\cite{Andreani/Bass/Pasq}.

The Hamilton operator \( H \), Eq.~(\ref{efmHam}), acts within these approximations
on a four-component envelope function in the product basis of the conduction
and valence band edge states 
\( \left\{ |\textstyle \frac{3}{2}\displaystyle \, m_{J}\rangle _{v}\, 
|\textstyle \frac{1}{2}\, +\frac{1}{2}\displaystyle \rangle _{c}\, \right\}  \),
where the hole-spin projection attains values of \( m_{J}=\textstyle +\frac{3}{2},+\frac{1}{2},-\frac{1}{2},-\frac{3}{2}\displaystyle  \):
\begin{equation}
\label{CB}
\begin{array}{rcl}
\displaystyle H_{e} & = & \left( -{\frac{\hbar ^{2}}{2m_{e}}}(\partial _{x_{e}}^{2}+\partial _{y_{e}}^{2}+\partial _{z_{e}}^{2})+V_{c}\right) \, I\,\,,
\end{array}
\end{equation}
 
\begin{equation}
\label{Lut}
\begin{array}{rcl}
\displaystyle H_{h} & = & -{\frac{\displaystyle \hbar ^{2}}{\displaystyle 2m_{0}}}\textstyle \left( \begin{array}{cccc}
\cal {P}+{\cal Q} & \cal {L} & \cal {M} & 0\\
\cal {L}^{\dagger } & \cal {P}-{\cal Q} & 0 & \cal {M}\\
\cal {M}^{\dagger } & 0 & \cal {P}-{\cal Q} & -\cal {L}\\
0 & \cal {M}^{\dagger } & -\cal {L}^{\dagger } & \cal {P}+{\cal Q}
\end{array}\right) +V_{v}\, I
\end{array}
\end{equation}
with 
\[
\begin{array}{lclcrcl}
\displaystyle \cal {P}\! \!  & =\! \!  & \gamma _{1}(\partial _{x_{h}}^{2}+\partial _{y_{h}}^{2}+\partial _{z_{h}}^{2}) & \hfill  & {\cal Q}\! \!  & = & \! \! \gamma _{2}\left( \partial _{x_{h}}^{2}+\partial _{y_{h}}^{2}-2\partial _{z_{h}}^{2}\right) \vspace {0.15cm}\\
\cal {L}\! \!  & =\! \!  & -i\, 2\sqrt{3}\gamma _{3}\, \left( \partial _{x_{h}}-i\partial _{y_{h}}\right) \, \partial _{z_{h}} & \hfill  & \cal {M}\! \!  & = & \! \! \sqrt{3}\, {\frac{\gamma _{2}+\gamma _{3}}{2}}\, (\, \partial _{x_{h}}^{2}-\partial _{y_{h}}^{2}-i\, 2\partial _{x_{h}y_{h}}^{2})\, 
\end{array}\]
and 
\begin{equation}
\label{Coulomb}
\begin{array}{ccl}
\displaystyle V_{Coul}(\vec{r}_{e}-\vec{r}_{h}) & = & -{{\frac{\displaystyle e^{2}}{\displaystyle \vphantom {M}\epsilon }}\frac{\displaystyle 1}{\displaystyle \mid \vec{r}_{e}-\vec{r}_{h}\mid ^{\vphantom {b}}}}\, I\,\,.
\end{array}
\end{equation}
 \( I \) is the \( 4\times 4 \) unity matrix. The material parameters \( \gamma _{1}(z_{h}),\gamma _{2}(z_{h}),\gamma _{3}(z_{h}) \)
as well as the offsets \( V_{v}(z_{h}),\, V_{c}(z_{e}) \) are piecewise constant
functions of \( z_{e},\, z_{h} \). To ensure that the kinetic operators remain
Hermitian in the presence of interfaces, we use the symmetric substitutions
\begin{equation}
\label{symmetric substitutions}
\gamma \, \partial _{i}\rightarrow (\partial _{i}\, \gamma +\gamma \, \partial _{i})/2\,,\, \, \, \gamma \, \partial _{ij}^{2}\rightarrow (\partial _{i}\, \gamma \, \partial _{j}+\partial _{j}\, \gamma \, \partial _{i})/2 \,,\, \, \, \, i=x,y,z \,. 
\end{equation}

The in-plane COM~momentum \( \vec{Q}=-i\hbar (\vec{\nabla }_{e_{\parallel }}+\vec{\nabla }_{h_{\parallel }}) \)
is a constant of motion because the interaction term (\ref{Coulomb}) depends
only on the relative distance of the two particles. 
Reflection with respect to the central 
\( xy \) plane, \( \sigma _{xy} \), is also a symmetry element
for symmetric QW. Consequently, the exciton can be characterized  by the parity
\( P=\pm 1 \). Then, the wavefunction factorizes into
\begin{equation}
\label{psi}
\displaystyle \Psi ^{\vec{Q};\, Pa}(\vec{r}_{e},\vec{r}_{h})\, =\, \frac{e^{-i\vec{Q}\cdot \vec{R}}}{2\pi }\, \, \sum _{m_{J}}\, \Psi ^{\vec{Q};\, Pa}_{m_{J}}(z_{e},z_{h},\vec{\rho })\mid \textstyle \frac{3}{2}\displaystyle \, \, m_{J}\rangle _{v}\, \, \mid \textstyle \frac{1}{2}\, \frac{1}{2}\displaystyle \rangle _{c}\,\, ,
\end{equation}
 where \( \vec{\rho }=\vec{r}_{e_{\parallel }}-\vec{r}_{h_{\parallel }} \)
is the in-plane particle distance, \( \vec{R} \) the COM~space coordinate
canonically conjugate to \( \vec{Q} \), and \( a \) stands for the remaining
quantum numbers related to the relative motion of the exciton. 

The COM~space coordinate \( \vec{R} \) in (\ref{psi}) is not unambiguously
defined because of the anharmonic dispersions of the constituent particles \cite{Dressel}.
The COM~transformation must be linear in order to preserve the canonical commutation
relations of space and momentum operators, and it has in general the form 
\begin{equation}
\label{R}
\vec{R}=\boldsymbol {\beta }\, \vec{r}_{e_{\parallel }}+\left( \boldsymbol {1}-\boldsymbol {\beta }\right) \, \vec{r}_{h_{\parallel }}\, ,\hspace {0.3cm}\vec{k}=-i\hbar \left( \left( \boldsymbol {1}-\boldsymbol {\beta }\right) \, \vec{\nabla }_{e_{\parallel }}-\boldsymbol {\beta }\, \vec{\nabla }_{h_{\parallel }}\right)\, .
\end{equation}
In the parabolic case, the free parameter~\( \boldsymbol {\beta } \) is
taken as the scalar 
\begin{equation}
\label{betaParabolic}
\beta _{parab}=\frac{m_{e}}{m_{e}+m_{h}}
\end{equation}
 in order that relative and COM~motion completely decouple. For bulk excitons,
\( \boldsymbol {\beta } \) has been considered in the literature as a scalar,
a tensor in real space \cite{Altar}, or even a spinor \cite{Kanehisa}. We will
return later on to the problem of an appropriate choice for the COM~coordinate
\( \vec{R} \).

Taking into account the electron spin degeneracy, each exciton state is at least
fourfold degenerate. It can be shown in a similar way as has been done for the
hole subband states in Ref.~\cite{Andreani/Bass/Pasq} that the operator \( R_{\pi }T \),
with the rotation by \( \pi  \) about the \( z \) axis, \( R_{\pi } \), and
time-reversal, \( T \), transforms between the degenerate states of different
parity and opposite electron spin. If one combines this operator with the Pauli matrix
\( \sigma _{y}^{e} \), which flips only the electron spin, we have apart from
an overall phase: 
\begin{eqnarray}
\displaystyle \Psi ^{\vec{Q};\, -Pa}(\vec{r}_{e},\vec{r}_{h}) & = & \left( \sigma ^{e}_{y}\, R_{\pi }\, T\right) \, \Psi ^{\vec{Q};\, Pa}(\vec{r}_{e},\vec{r}_{h})\label{-P} \\
 & = & \frac{e^{-i\vec{Q}\cdot \vec{R}}}{2\pi }\, \, \sum _{m_{J}}\, \Psi ^{\vec{Q};\, Pa\, ^{*}}_{m_{J}}(z_{e},z_{h},-\vec{\rho }\, )\mid \textstyle \frac{3}{2}\displaystyle \, \, -m_{J}\rangle _{v}\, \, \mid \textstyle \frac{1}{2}\, +\frac{1}{2}\displaystyle \rangle _{c}\,\, .\nonumber 
\end{eqnarray}
Comparing (\ref{psi}) with (\ref{-P}), we find:
\begin{equation}
\label{-P2}
\Psi ^{\vec{Q};\, -Pa}_{m_{J}}(z_{e},z_{h},\vec{\rho }\, )=\Psi ^{\vec{Q};\, Pa\, ^{*}}_{-m_{J}}(z_{e},z_{h},-\vec{\rho }\, )\,\, .
\end{equation}
 That is, the (degenerate) state of reversed parity 
is obtained by inverting the order of the spin components of the exciton
envelope, complex conjugating, and changing the sign of the in-plane relative
coordinate. Thus changing the multiband exciton parity with fixed electron
spin in symmetric QW corresponds to flipping the hole spin in the single-band
exciton case. In the axial approximation and for \( Q=0 \), the different angular
momentum components decouple \cite{Zhu/Huang}, and changing the sign of \( \vec{\rho } \)
in Eq.~(\ref{-P2}) just changes the sign of two spin components leaving the
other two unchanged; this holds no longer at \( Q\neq 0 \).

We have solved the eigenvalue problem (\ref{efmHam}) in two ways which will be discussed
in turn:
(i) in \( \vec{k} \)-space, expanding Eq.~(\ref{psi}) in the product space
of the electron and hole subband states \cite{Broido/Sham}, and 
(ii) in real space, using a finite-differences scheme.
The first method gives very accurate results and is used to reveal the main
features of the exciton dispersion. The second method is only suitable for the
groundstate dispersion but promises to be feasible for more general structures.
It is validated by comparing its results with the ones from the first method.

\subsection{Solution in \protect\( \vec{k}\protect \)-space\label{SKSPACE}}

As a first step, we calculate the single-particle subband states and their dispersions
\begin{eqnarray}
\displaystyle H_{e}\, |n_{e}\, \vec{k}_{e};\textstyle \pm \frac{1}{2}\displaystyle \rangle  & = & {\cal E}_{n_{e}}(k_{e})\, |n_{e}\, \vec{k}_{e};\textstyle \pm \frac{1}{2}\displaystyle \rangle \nonumber \\
H_{h}\, |n_{h}\, \vec{k}_{h};\, p_{h}\rangle  & = & {\cal E}_{n_{h}}(k_{h})\, |n_{h}\, \vec{k}_{h};\, p_{h}\rangle \label{1Pdispersions} 
\end{eqnarray}
 using a transfer-matrix method as in \cite{RamMohan}. The respective solutions
in the axial approximation are of the form
\begin{eqnarray}
\displaystyle |n_{e}\, \vec{k}_{e};\textstyle \pm \frac{1}{2}\displaystyle \rangle  & = & \frac{e^{i\vec{k}_{e}\cdot \vec{r}_{e_{\parallel }}}}{2\pi }\, e^{i\left( \scriptstyle \pm \textstyle \frac{1}{2}\right) \theta _{e}}\xi _{n_{e}}(z_{e})\, |\textstyle \frac{1}{2}\, \pm \frac{1}{2}\displaystyle \rangle _{c}\,\, ,\label{1P} \\
|n_{h}\, \vec{k}_{h};\, p_{h}=\pm 1\rangle  & = & \frac{e^{i\vec{k}_{h}\cdot \vec{r}_{h_{\parallel }}}}{2\pi }\, \sum _{m_{J}}\, e^{i\, m_{J}\, \theta _{h}}\, \xi ^{m_{J}}_{n_{h},\, p_{h},\, k_{h}}(z_{h})\, |\textstyle \frac{3}{2}\displaystyle \, m_{J}\rangle _{v}\,\, .\nonumber 
\end{eqnarray}
 In Eqs.~(\ref{1Pdispersions},\ref{1P}) \( n_{e,h} \) denote the subband indices,
\( \vec{k}_{e,h}=\left( k_{e,h},\, \theta _{e,h}\right)  \) the respective
in-plane wavevectors in polar coordinates and \( p_{h} \) the hole parity under
\( \sigma _{xy} \) \cite{Andreani/Bass/Pasq}. 

In a second step, the exciton wavefunction for a given COM~momentum \( \vec{Q} \)
is expanded into 
\begin{eqnarray}
\displaystyle \Psi ^{\vec{Q};\, P a }\left( \vec{r}_{e},\, \vec{r}_{h}\right)  & \! \! = & \! \! \! \sum _{n_{e}n_{h}}\! \int d\vec{k}\, \, \varphi _{n_{e}n_{h}}^{\vec{Q};\, a}(\vec{k})\, |n_{e}\, \vec{k}_{e}\, \textstyle ;+\frac{1}{2}\displaystyle \rangle _{c}\, |n_{h}\, \vec{k}_{h};\, p_{h}\rangle _{v}\,\, ,\label{exciton1} 
\end{eqnarray}
with subband states of the two particles combined in such a way that the resulting
exciton state has the required parity \( P \) and total momentum \( \vec{Q} \):
\begin{equation}
\label{exciton2}
\vec{k}_{e}=\vec{k}+\boldsymbol {\beta }\vec{Q} \,,\hspace {0.5cm}\, \vec{k}_{h}=\vec{k}-\left( \boldsymbol {1}-\boldsymbol {\beta }\right) \vec{Q}\, ,\hspace {0.5cm}\, P=p_{h}\cdot (-1)^{n_{e}+1}\, .
\end{equation}
The last relation reflects that the conduction subband envelopes are even (odd)
for odd (even) subband index. Fixing exciton parity \( P \) and electron spin
eliminates any degeneracy at \( \vec{Q}\neq 0 \). 

With the expansion (\ref{exciton1}) and the relations (\ref{exciton2}), the
exciton Schr\"odinger equation takes the form 
\begin{eqnarray}
\left( {\cal E}_{n_{e}}(\vec{k}_{e})+{\cal E}_{n_{h}}(\vec{k}_{h})-E_{a}^{X}(\vec{Q}\, \! )\right) \, \varphi _{n_{e}n_{h}}^{\vec{Q};\, a}(\vec{k}\, \! )\, + \sum _{n_{e}'n_{h}'}\int \! d\vec{k}'\, V^{\vec{Q}}_{n_{e}'n_{h}'\atop n_{e}n_{h}}(\vec{k},\vec{k}')\, \varphi _{n_{e}'n_{h}'}^{\vec{Q};\, a}(\vec{k}')=0\,\, ,\label{Hk}
\end{eqnarray}
where \( E_{a}^{X}(\vec{Q}\, \! ) \) denotes the energy dispersion of the exciton
state. The interaction 
\begin{equation}
\label{kCoulomb}
V^{\vec{Q}}_{n_{e}'n_{h}'\atop n_{e}n_{h}}(\vec{k},\vec{k}')=-\frac{1}{2\pi }\frac{e^{2}}{\epsilon }\, \frac{1}{|\vec{k}-\vec{k}'|}\, {\cal F}_{n_{e}'n_{h}'\atop n_{e}n_{h}}^{\, \vec{Q}}(\vec{k},\vec{k}')
\end{equation}
is the in-plane 2D Fourier transform of the 3D Coulomb potential modified due
to the confinement in \( z \)-direction. The latter is expressed through form
factors [with Eq.~(\ref{exciton2})]
\begin{eqnarray}
{\cal F}_{n_{e}'n_{h}'\atop n_{e}n_{h}}^{\, \vec{Q}}(\vec{k},\vec{k}') & = & \sum _{m_{J}}\iint dz_{e}\, dz_{h}\, e^{-|\vec{k}-\vec{k}'|\, |z_{e}-z_{h}|}\,   \xi _{n_{e}}^{*}(z_{e})\, \xi _{n_{h},\, \vec{k}_{h}}^{m_{J}\, ^{*}}(z_{h})\, \xi _{n_{e}'}(z_{e})\, \xi ^{m_{J}}_{n_{h}',\, \vec{k}_{h}'}(z_{h})\,\, . \label{formfactors}
\end{eqnarray}
 The above integrals are calculated analytically since the subband states obtained
with the transfer matrix method are combinations of exponential and trigonometric
functions. 

The integrable singularity of the Coulomb potential (\ref{kCoulomb}) at \( \vec{k}=\vec{k}' \)
is taken care of by adding and subtracting in Eq.~(\ref{Hk}) the analytically
integrable term 
\begin{equation}
\label{sC}
C(\vec{k},\, \vec{k}')=\frac{e^{2}}{\epsilon }\, \frac{1}{2\pi }\left( \frac{1}{|\vec{k}-\vec{k}'|}-\frac{1}{\vphantom {B}max\left( k,\, k')\right) }\right) \, .
\end{equation}
 This gives a smooth ``corrected'' potential (\ref{kCoulomb}) of small  
absolute magnitude.

To take benefit of the axial approximation, the exciton envelope and the form
factors are expanded into 2D angular momentum eigenstates \( \exp({i\ell \theta }) \).
The angular momentum \( \ell  \) is chosen for every subband combination
such that \( \ell \! =\! 0 \) corresponds to the respective \( s \)-like exciton
\cite{Choa/Chuang} at \( Q=0 \).

 The resulting set of coupled one-dimensional integral equations
is solved numerically for various values of the COM~momentum \( Q \). Results
will be presented in section \ref{RESULTS}.

\subsection{Solution in real space\label{REALSPACE}}

We demonstrated in Ref.~\cite{US} that calculations of the 
multiband-exciton groundstate dispersion
are also feasible with a finite-differences scheme in real space.
This method is conceptually simple: 
the Schr\"odinger equation corresponding to 
Eqs.~(\ref{efmHam})-(\ref{Coulomb}),(\ref{psi})
leads to a system of four (number of spin components) coupled partial differential
equations in the four dimensional space \( (\vec{\rho },z_{e},z_{h}) \). 
The resulting eigenvalue problem involves a 
large sparse complex Hermitian matrix with a substantial
number, 44, of nonzero off-diagonals. In contrast to the \( \vec{k} \)-space
approach, the method can, in principle, be applied effortlessly to very general
heterostructures, like quantum wires and quantum dots. The main drawback is
the need for huge amounts of computer memory. Indeed, the dimension of the matrix
to be diagonalized scales with the fourth power of the number of grid points
per spatial dimension. The most dense grid we used led to a matrix of dimension
\( N_{D}=4\times 31\times 31\times 61\times 61=1.4\cdot 10^{7} \) (4 is the
number of spin components). 

We used the ARPACK \cite{ARPACK} package to calculate a few eigenvalues and
eigenvectors at the lower end of the spectrum. ARPACK is an efficient implementation
\cite{Schreiber} of the Implicitly Restarted Arnoldi Method that can be viewed
as a synthesis of the Arnoldi/Lanczos process with the Implicitly Shifted QR
algorithm. Storage of the (nonzero) matrix elements is not required, only a
matrix-vector multiplication utility is needed. Multiple eigenvalues, as they
occur in our case, offer no additional problems. However, if one needs all the
members of a multiplet the iteration subspace has to be chosen large enough.
We find that an iteration subspace of five times the number of the requested
eigenvalues (rather than the proposed factor of two \cite{ARPACK}) is usually
sufficient. This matter was of no concern for our problem, since using the symmetry
considerations, Eq.~(\ref{-P2}), we can derive from a calculated state also
the second one of the doublet. For the largest matrices, we used a factor of
three as a compromise between memory demand and CPU time usage. 

The matrix resulting from the discretization is highly structured. For minimizing
the memory costs and still making full use of the vector registers, we construct
the matrix-vector product using auxiliary, much smaller matrices. 

Since memory is critical, it is crucial for any real-space approach to optimize
the convergence of the relevant quantities with the mesh size. On the one hand,
we optimize the COM~transformation, as will be discussed in the next section,
thereby improving the handling of the kinetic terms. On the other hand, we use
a groundstate-adapted discretization of the Coulomb potential \cite{Glutsch}
which is discussed in more detail in the Appendix. 
The idea behind this approach is to extract the discretized interaction 
from a reference
system that has the same interaction but a simple kinetic term, 
and whose groundstate is known analytically.
If the groundstate of the reference system is similar enough to the one of
the real system, good convergence is expected.

Calculations on a parallel-vector machine of type CRAYJ932 reached performances
of 140MFlops/CPU, the peak performance of the limiting BLAS routine being 185MFlops/CPU.

\section{Optimized center-of-mass transformation\label{CMM}}

We return to the ambiguity in the COM~transformation (\ref{R}) which is
expressed in the freedom to choose~\( \boldsymbol {\beta } \). The
relevance of \( \boldsymbol {\beta } \) for accelerating numerical convergence
in dispersion calculations was realized quite early for bulk excitons in \cite{Altar},
where a scalar \( {\beta } \) was optimized in a trial and error
procedure. However, there has been no other algorithm to take advantage of this
freedom until recently \cite{US}. Before that, there have been just two publications
where numerical multiband exciton dispersion in quantum wells were calculated:
in Ref.~\cite{JorRoesBro} \( \beta =1 \) (in the parabolic case \( m_{e}=\infty  \))
was taken in order for the form factors (\ref{formfactors}) to be independent
of \( Q \), and in Ref.~\cite{Triques} no particular choice or handling of
\( \beta  \) is mentioned. In analytic expressions, usually the symmetric (in
the parabolic case \( m_{e}=m_{h}) \) value \( \beta =1/2 \) is taken \cite{Winkler}.

\begin{figure}
\begin{center}
\resizebox*{6.5cm}{!}{\includegraphics{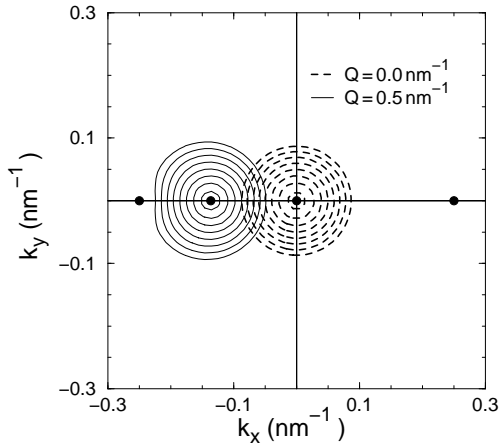}}
\end{center}
\caption{\label{Fig_SHIFT_k}Logarithmic contour plots of the 
squared \protect\( HH_{1}C_{1}-1s\protect \)
exciton envelope in \protect\( \vec{k}\protect \)-space for a 5~nm GaAs/Al\protect\( _{0.3}\protect \)Ga\protect\( _{0.7}\protect \)As
QW at \protect\( Q_{x}=0.5\protect \)~nm\protect\( ^{-1}\protect \) plotted
in a coordinate system with the not optimized value 
\protect\( \beta =1/2\protect \). 
Diamonds mark, from left to right,
the origin
of the shifted coordinate systems [at  \protect\( (1/2-\beta) Q_{x}\protect \)]
for \protect\( \beta =0,\, \beta _{0},\, 1/2,\, 1\protect \),
with \protect\( \beta _{0}\protect \) evaluated from Eq.~(\ref{betaSAext}).
The \protect\( Q=0\protect \) envelope is plotted for comparison (dashed).}
\end{figure}

The effect of the \( \beta  \) choice becomes clear when one evaluates Eq.~(\ref{R})
for two different values 
\( \beta ,\, \beta '=\beta +\delta {\beta}  \) giving
\( \vec{R}'=\vec{R}+(\delta {\beta}) \, \vec{\rho },\, \, \vec{k}'
 =\vec{k}-(\delta {\beta}) \, \vec{Q} \).
Clearly, \( \beta  \) moves artificially part of the plane wave of the COM~motion
into the relative part of the exciton (\ref{psi}) or, equivalently, it shifts
the relative part of the wavefunction in \( \vec{k} \)-space. A good choice
of \( \beta  \), as in the parabolic case (\ref{betaParabolic}), keeps the
relative part of the exciton in real space as smooth as possible or, equivalently,
pins the relative part of the wavefunction in \( \vec{k} \)-space to the origin.
This situation is illustrated in Fig.~\ref{Fig_SHIFT_k} where we plot the envelope
of the \( HH_{1}C_{1}-1s \) exciton in the single subband approximation using
the symmetric value of \( \beta =1/2 \). This value of \( \beta  \) is indeed
not optimal, as the large shift demonstrates. Diamonds mark where the position
of the origin would be for other values of \( \beta  \).

We introduced in Ref.~\cite{US} a quasi-analytical method for determining the
optimal choice of the scalar \( \beta  \), which we briefly summarize here.
It is motivated by the fact that in the parabolic case the correct COM~transformation
decouples the relative motion and COM~motion completely. 
A full decoupling
is not possible for non-parabolic dispersions. We looked for a choice of \( \beta  \)
that decouples ``as much as possible''. To quantify this, we inserted in Eq.~(\ref{efmHam})
the general \( \beta  \)-dependent COM transformation (\ref{R}), separated
the \( Q \)-dependent terms from the rest 
\begin{equation}
\label{pert}
\begin{array}{lcl}
H & = & H^{(0)}+H^{(1)}(\beta )\, Q+H^{(2)}(\beta )\, Q^{2}
\end{array}\, ,
\end{equation}
 and viewed these as a perturbation of the \( Q=0 \) exciton. Taking into account
the inversion symmetry of the Brillouin zone, the kinetic mass of the groundstate
exciton \( g \) is given in second order perturbation theory by
\begin{equation}
\label{pert2}
\frac{\hbar ^{2}}{2M_{g}^{X}}=\langle \, g\mid H^{(2)}(\beta )\mid g\, \rangle +\sum _{a\neq g}\frac{|\, \langle a\mid H^{(1)}(\beta )\mid g\rangle \, |^{2}}{E_{g}^{X}(0)-E^{X}_{a}(0)}\,\,.
\end{equation}
However, the exciton mass must not depend on \( \beta  \). Maximizing the
first order contribution in (\ref{pert2}) and minimizing this way the strictly
positive contribution of the higher states to the groundstate mass leads to
the analytical result 
\begin{equation}
\label{betaOld}
\begin{array}{lcl}
\beta  & = & \langle \, g\mid H^{(2)}_{h}|\, g\, \rangle \, 
 \left / \,
 \langle \, g\mid H^{(2)}_{e}+H^{(2)}_{h}|\, g \, \rangle 
 \right.
\end{array}\, .
\end{equation}
 \( H_{e,h}^{(2)} \) are simply the material-dependent coefficients of the
$\beta^2 Q^2$ terms 
when inserting (\ref{exciton2}) into the $k$-space representation of 
the kinetic energies in Eq.~(\ref{CB})
and (\ref{Lut}), respectively \cite{US}. 

The explicit form of Eq.~(\ref{pert2}) with the contributions from the higher
states dropped and \( \beta  \) from Eq.~(\ref{betaOld}) suggests to define
COM-related, effective masses \( m^{*}_{e,h} \) for electron and hole 
\begin{equation}
\label{meffOld}
1/m^{*}_{e,h}=\frac{2}{\hbar ^{2}}\,\langle g\, | H^{(2)}_{e,h}|\, g\rangle 
\, \, \, \mbox {~~satisfying}\hspace {0.5cm}M_{g}^{X}=m_{e}^{*}+m_{h}^{*}\,.
\end{equation}
 Numerical results show that the masses obtained from (\ref{meffOld}) tend
to be too small. Nevertheless, the obtained values for \( \beta  \) in 
\cite{US,US2} were quite reasonable because of the much heavier hole mass. If
one actually calculates the contributions of the higher exciton states to \( M_{g}^{X} \)
in (\ref{pert2}) (which are dropped in (\ref{meffOld})), one finds that the
only important correction comes from the coupling to the \( LH_{1}C_{1}-1s \)-like
state. Taking in Eq.~(\ref{pert2}) this single correction term into account
gives practically the exact curvature of the exciton groundstate dispersion
at \( Q=0 \).

The above procedure is not the best for determining the optimal value of \( \beta  \)
as the importance of the coupling to higher states demonstrates. It was inspired
by the solution of the exciton problem in real space. Let us now look at the
form of the exciton wavefunction for \( Q\neq 0 \) in the subband expansion
(\ref{exciton1}). The \( Q \)-dependence enters the wavefunction: 
(i)
the need to appropriately combine the subband states to get the right \( Q \)
and 
(ii) through the need for the envelope to adjust for the anharmonicities
in the dispersions. In the perturbation approach described above, we tried to
find a COM~transformation that keeps 
for small \( Q \) values the \emph{entire wavefunction} 
unchanged as much
as possible. However, once the one-particle
problem is solved, the \( Q \)-dependence due to the appropriate combination
of the subband states (i) is explicitly known. Therefore, a better Ansatz for
the wavefunction would be to find a COM~transformation that keeps
the \emph{envelopes} as much unchanged as possible; that is Eq.~(\ref{exciton1})
with
\begin{equation}
\label{shiftA}
\varphi _{n_{e}n_{h}}^{\vec{Q};\, g}(\vec{k})=\varphi _{n_{e}n_{h}}^{0;\, g}(\vec{k})\,\, .
\end{equation}
The minimization of the energy  with respect to \( \beta  \) can
be done analytically in the limit \( Q\, \rightarrow \, 0 \). We easily obtain
the optimized value 
\begin{equation}
\label{betaSAext}
\beta _{0}=\frac{\displaystyle \sum _{n_{e}n_{h}}\, \int \! d\vec{k}\, |\, \varphi ^{0;\, g}_{n_{e}n_{h}}(\vec{k})\, |^{2}\, (\hat{Q}\cdot \vec{\nabla })^{2}{\cal E}_{n_{h}}(\vec{k})}{\displaystyle \sum _{n_{e}n_{h}}\, \int \! d\vec{k}\, |\, \varphi ^{0;\, g}_{n_{e}n_{h}}(\vec{k})\, |^{2}\, (\hat{Q}\cdot \vec{\nabla })^{2}({\cal E}_{n_{e}}(\vec{k})+{\cal E}_{n_{h}}(\vec{k}))}\,\, .
\end{equation}
 This expression accounts also for the dependence of \( \beta _{0} \) on the
direction \( \hat{Q} \) of the COM momentum in the case of warped valence
bands. 
We took into account the inversion symmetry of the Brillouin zone and assumed
that the Coulomb potential (\ref{kCoulomb}), (\ref{formfactors}) can be approximated
as a function of the momentum transfer only, \( V(\vec{k},\, \vec{k}')\simeq V(\vec{k}-\vec{k}') \),
i.e., we neglected any \( Q \)-dependence of the (Coulomb) potential energy
of the groundstate. This is expected to be a good approximation since the Coulomb
energy depends solely on the charge distribution, which should not be affected
significantly by the in-plane motion. Indeed, it has been estimated in Ref.~\cite{Sander/Chang}
that the error introduced by neglecting in Eq.~(\ref{formfactors}) the \( \vec{k} \)-dependence
of the hole envelopes is about \( 5\% \). Our assumption should lead to even
smaller deviations.

Fig.~\ref{Fig_SHIFT_k} demonstrates the quality of the expression~(\ref{betaSAext}). 
It shows
the groundstate envelope at a rather large value of \( Q=0.5 \)~nm\( ^{-1} \),
even though Eq.~(\ref{betaSAext}) was obtained in the limit \( Q \rightarrow 0 \).
It is particularly impressive that the respective origin 
for the optimized choice~\( \beta _{0} \) 
from Eq.~(\ref{betaSAext}) lies even a bit to the left of the maximum
of the envelope. This accounts for the \( p \)-component that deforms the originally
radially symmetric \( Q=0 \) envelope; using an angular momentum decomposition
of the envelope a minimum number of components would be needed. The slight deformation
(these are logarithmic plots) for large values of \( Q \) is due to the anharmonicity
of the one-particle dispersions.

\begin{figure}
\begin{center}
\resizebox*{8.5cm}{!}{\includegraphics{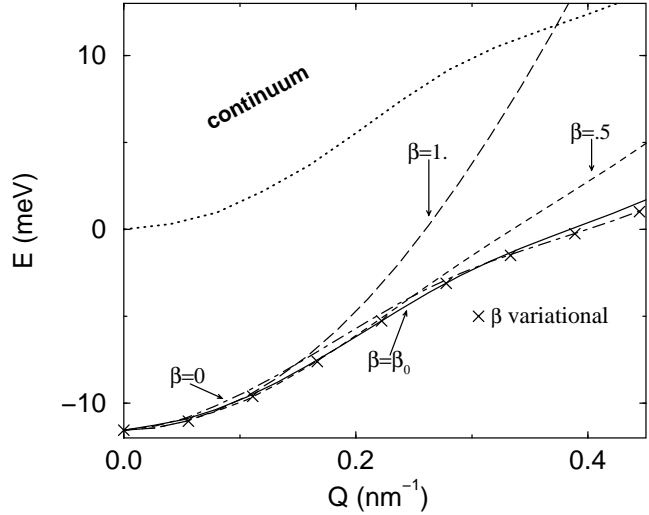}}
\end{center}

\caption{\label{Fig_BetaImp}Dispersion of the groundstate exciton of a 5~nm wide QW
calculated with the same basis but for various values of \protect\( \beta \protect \).
The exciton continuum edge (dotted), Eq.~(\ref{XContEdge}), is given for comparison.}
\end{figure}

The importance of a suitable choice of the COM~transformation for the numerical
convergence is illustrated in Fig.~\ref{Fig_BetaImp} for the dispersion of
the \( HH_{1}C_{1}-1s \) exciton of a 5~nm QW. This has been calculated in \( \vec{k} \)-space
for various values of \( \beta  \) with the same basis (\( HH_{1}C_{1},\, LH_{1}C_{1},\, \ell =0,\pm 1,\pm 2 \)).
The further the used \( \beta  \) lies from the optimal value \( \beta _{0} \)
(\( \beta _{0}=0.23 \) in this case) the worse the results are. We did also
calculations where for \emph{each} \( Q \) value an optimal value of \( \beta  \)
was obtained by numerical variation. We observed deviations from \( \beta _{0} \)
less than \( 1\% \) near \( Q=0 \) and not larger than \( 10\% \) at \( Q=0.5 \)~nm\( ^{-1} \)
even for the widest well (20~nm). At large \( Q \) the \( \beta =0 \) (in the
parabolic case \( m_{h}=\infty  \)) curve gives slightly better results than
\( \beta _{0} \) since the \( HH_{1} \) subband dispersion gets more flat
after the avoided crossing with the \( LH_{1} \) subband, but it gives considerably
worse results at small \( Q \).

\section{A simple analytical formula for the average exciton kinetic mass\label{SMASS}}

The analytical variation that led to Eq.~(\ref{betaSAext}) gives the groundstate
energy up to terms quadratic in \( Q \). The corresponding groundstate kinetic
mass \( M_{g}^{X} \) has again the form
\begin{equation}
\label{newMX}
M_{g}^{X}=m^{X}_{e}+m^{X}_{h}
\end{equation}
with the COM-related effective masses for electron and hole defined as 
\begin{equation}
\label{effmassSA}
1/m^{X}_{e,h}=\frac{1}{\hbar ^{2}}\, \sum _{n_{e}n_{h}}\, \int d\vec{k}\, |\, \varphi ^{0;\, g}_{n_{e}n_{h}}(\vec{k})\, |^{2}(\hat{Q}\cdot \vec{\nabla })^{2}{\cal E}_{e,h}(\vec{k})\,\,.
\end{equation}
With these masses, the expression for \( \beta _{0} \), Eq.~(\ref{betaSAext}), has
the same form as in the parabolic case (\ref{betaParabolic}). Eq.~(\ref{effmassSA})
gives the correct results for the free particle case. 

We claim that this simple result will be of considerable practical importance. Eq.~(\ref{effmassSA})
is physically appealing: it leads to a weighted average of the subband dispersions.
Further, it is relatively simple
to calculate: It requires only the approximate knowledge of the  \( Q=0 \) exciton
 envelope 
 and of the involved subband dispersions. The numerical
calculation of subband dispersions is nowadays an easy task (provided the \( \vec{k}\cdot \vec{p} \)
parameters are known). Moreover, especially for narrow QW, the envelope of the
\( HH_{1}C_{1} \) component of the groundstate exciton is to a very good approximation
similar in shape to the groundstate of the 2D exciton, 
\begin{equation}
\label{2D-1s}
\varphi ^{2D}_{1s}(\vec{\rho }\, )=\sqrt{\frac{2}{\pi a_{B}^{2}}\, } \, e^{-a_{B}\, \rho }
  ~,~~
\, \varphi _{1s}^{2D}(\vec{k}\, )= \sqrt{\frac{2a_{B}^{2}}{\pi }}\,  \left( 1 + \left( a_{B}k\right) ^{2}\right) ^{-3/2}.
\end{equation}
 The \( LH_{1}C_{1} \) component is quite small, e.g.,~\( 5\% \) for the 20~nm
QW, and can be safely neglected in this context. Therefore, only a good estimate
for the effective Bohr radius \( a_{B} \) is needed to evaluate (\ref{effmassSA}).

\section{Results\label{RESULTS}}

We have calculated exciton dispersions both in real and momentum space for GaAs/\-Al\( _{0.3} \)Ga\( _{0.7} \)As
\( \langle 001\rangle  \) QW of various widths. The coupling of heavy and light
holes was fully incorporated. 
The values of the material parameters
 \( \gamma _{1},\, \gamma _{2},\, \gamma _{3},\, m_{e} \)
were taken by linear interpolation from the GaAs and AlAs values; 
the offset ratio was \( V_{v}/V_{c}=0.68/0.32 \) and the band gap in meV was 
taken as \( E_{g}(x)=1519+1040\, x+470\, x^{2} \),
\( x \) being the Al content\cite{Landolt-Bornstein}. 
For the dielectric constant, we adopted  \( \epsilon =12 \) for both well 
and barrier material.

\subsection{Subband expansion\label{KSECTION}}

The nomenclature is as follows: the exciton in the subband
expansion has various \( n_{h}n_{e} \) subband components with the corresponding
envelopes \( \varphi ^{\vec{Q};\, a}_{n_{e}n_{h}}(\vec{k}) \), Eq.~(\ref{exciton1}).
These envelopes have in the axial approximation at \( Q=0 \) a definite angular
momentum \( \ell  \) and will be denoted by \( 1s,\, 2s,\, 2p_{\pm },\, 3d_{\pm } \)
and so on. Each exciton state at finite \( Q \) will be named according to
the main subband component of the corresponding state at \( Q=0 \). That is,
speaking of the \( HH_{1}C_{1}-1s \) exciton means that at \( Q=0 \)
its main subband component is the \( HH_{1}C_{1} \) one with an \( 1s \) envelope.
Similar to the single-particle hole subband states, which can change their heavy-
or light-hole character away from the \( \Gamma  \)-point, the envelope of
the main subband component or even the main subband component itself can change
with increasing \( Q \). To denote the main subband component of a state at
a given value of \( \vec Q \), 
we will speak of the character of the state at this \( \vec Q \).
For example, 
 the \( HH_{1}C_{1}-2p_{+} \) exciton has a \( HH_{1}C_{1}-2p_{+} \) character
at \( Q=0 \) and a  \( LH_{1}C_{1}-1s \) character 
 \( Q \gg  0 \).

The exciton dispersions in \( k \)-space are calculated as follows: For each
QW, we first calculate the exciton spectrum at \( Q=0 \). Subsequently, a 2D
\(1s\)-exciton groundstate function (\ref{2D-1s}) is fitted to the \( HH_{1}C_{1} \)
envelope. 
For the wider QW, also a two-dimensional $3d$-exciton function 
is fitted to the \( LH_{1}C_{1} \)
envelope. This fit is used, instead of the numerical envelope, to evaluate the
optimized COM~transformation (\ref{betaSAext}) because it allows to take
advantage of the analytically known derivatives of the fit function. The so
calculated value of \( \beta _{0} \) is used for the \( Q\neq 0 \) calculations.

\begin{figure}
\begin{center}
\resizebox*{8.5cm}{!}{\rotatebox{-90}{\includegraphics{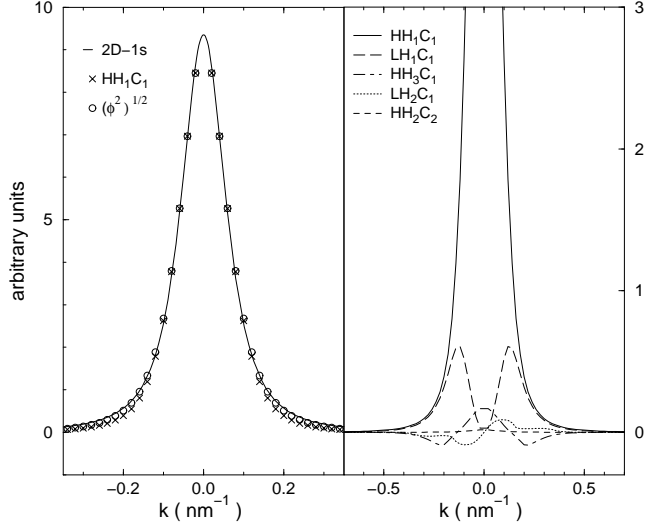}}}
\end{center}

\caption{\label{Fig_k-envelopes}On the right: Envelopes of the subband components of
the groundstate exciton calculated in \protect\( k\protect \)-space for a 15~nm
wide QW at \protect\( Q=0\protect \). On the left: \protect\( HH_{1}C_{1}\protect \)
envelope (\protect\( \times \protect \)), fitted 2D \protect\(1s\protect\)-exciton wavefunction
(line), and square root of the total probability distribution 
\protect\( |\phi(k)|^{\,2}=\sum _{n_{e}n_{h}}\varphi _{n_{e}n_{h}}^{0;\, g}(k)^{2}\protect \)
(circles).}
\end{figure}

In Fig.~\ref{Fig_k-envelopes}, we display the envelopes of the components of
the groundstate exciton in the subband expansion (\ref{exciton1}) for a 15~nm
wide QW at \( Q=0 \). The coupling of the higher subbands is rather small,
less than \( 3\% \) for the \( LH_{1}C_{1} \) component and even less for
the others. The parity and spin selection rules for the Coulomb coupling of
the subbands at the \( \Gamma  \)-point in symmetric QW are obeyed, e.g., the
admixture of the \( LH_{1}C_{1} \) state vanishes at the \( \Gamma  \)-point
since the Coulomb potential is spin-diagonal and the \( HH_{1} \) and \( LH_{1} \)
subband states are pure heavy- and light-hole states, respectively. The \( HH_{1}C_{1} \)
envelope is very well approximated by a 2D \(1s\)-exciton function.
Deviations are mainly located at the vicinity of the \( HH_{1}-LH_{1} \) avoided
crossing of the hole-subband dispersions 
(here, \( k_{ac}\simeq 0.13\)~nm\( ^{-1} \)).
The total in-plane probability distribution follows the form of Eq.~(\ref{2D-1s})
even better than the \( HH_{1}C_{1} \) envelope alone; the coupling to the
higher subbands allows the exciton to relax further. This, again, supports the
notion that the subband mixing has little influence on the charge distribution.

The calculations in \( k \)-space presented in the following take into account
only the two lowest hole subbands (\( n_{h}n_{e}=HH_{1}C_{1}, \) \( LH_{1}C_{1} \)).
Inclusion of higher subbands does not enhance the binding energy of the groundstate
exciton considerably. For the angular decomposition of the envelope components
only the \( s,\, p_{\pm },\, d_{+} \) (\( \ell =0,\pm 1,2) \) components for
the \( HH_{1}C_{1} \) and the \( s,\, p_{\pm },\, d_{-} \) (\( \ell =0,\pm 1,-2) \)
components for the \( LH_{1}C_{1} \) were considered. 
 Due to the optimized choice of the COM~coordinate
system, Eq.~(\ref{betaSAext}), these few angular momentum components are sufficient
to describe the \( HH_{1}C_{1}-1s \) and \( LH_{1}C_{1}-1s \) dispersions
excellently over the whole range of COM~momentum values considered, 
\( Q\leq 0.5 \)~nm\( ^{-1} \): 
The \( p \) components account mainly for the deformation of the
envelope, and the \( d \) components take care of the Coulomb coupling to higher
states.
\begin{figure}
\begin{center}
\resizebox*{8.5cm}{!}{\includegraphics{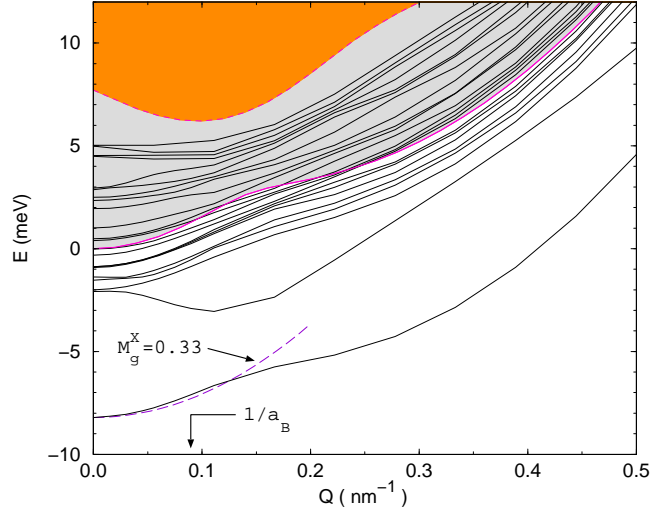}}
\end{center}

\caption{\label{Fig_k-disp}Exciton dispersions, groundstate (thick) and some excited
bound and continuum states (thin), calculated in \protect\( k\protect \)-space
for a 15~nm wide QW. The light (dark) gray shaded area
mark the \protect\( HH_{1}C_{1}\protect \) (\protect\( LH_{1}C_{1}\protect \))
exciton continuum. A parabola for the average groundstate exciton mass \protect\( M_{g}^{X}\protect \)
after Eqs.~(\ref{newMX},\ref{effmassSA}) is plotted (dashed). An arrow marks the
position of \protect\( 1/a_{B}\protect \) (\protect\( a_{B}\protect \) the
Bohr radius).}
\end{figure}

In Fig.~\ref{Fig_k-disp}, we show the dispersion calculated in \( k \)-space
of the first bound exciton states as well as some of the continuum states in
a 15~nm  wide QW. Zero of energy is the onset of the \( HH_{1}C_{1} \)
continuum at \( Q\! =\! 0 \). The Bohr radius \( a_{B} \) is the one obtained
from the same fit of a 2D \(1s\)-exciton to the envelope of the \( HH_{1}C_{1} \)
component at \( Q=0 \) 
which was used to calculate the optimized \( \beta _{0} \).
The groundstate is \( HH_{1}C_{1}-1s \). The next state is \( HH_{1}C_{1}-2p_{+} \)
which turns into  \( LH_{1}C_{1}-1s \) character at the avoided crossing.
The \( LH_{1}C_{1}-1s \) exciton is at \( Q=0 \) the fourth excited state
and shows a substantial mixing with the \( HH_{1}C_{1}-3d_{+} \) exciton. The
\( HH_{1}C_{1}-2p_{+} \) exciton has at \( Q=0 \) a slightly lower energy
than the \( HH_{1}C_{1}-2p_{-} \) (third state at \( Q\! =\! 0 \)) because it
couples to the \( LH_{1}C_{1}-1s \) outside of the \( \Gamma  \)-point. 

\begin{figure}
\begin{center}
\resizebox*{8.5cm}{!}{\includegraphics{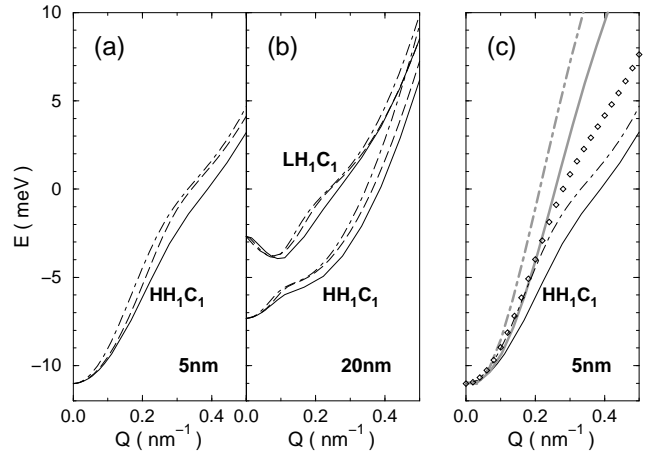}}
\end{center}

\caption{\label{Fig_CONT_Contra_DISP} (a), (b): Comparison of the dispersion of \protect\( HH_{1}C_{1}-1s\protect \)
and \protect\( LH_{1}C_{1}-1s\protect \) (20~nm QW only) excitons (thick) with
the appropriately shifted respective exciton continuum edge (dashed) and the
underlying hole-subband dispersion (dot-dashed) for two GaAs/Al\protect\( _{0.3}\protect \)Ga\protect\( _{0.7}\protect \)As
QW. (c): Comparison of our results (black) with those of Ref.~\protect\cite{Triques}
(gray) for the 5~nm QW (shifted to match at \protect\( Q=0\protect \)). The
exact \protect\( HH_{1}\protect \)-subband dispersion for the parameters of
\protect\cite{Triques} is plotted, too (diamonds).}
\end{figure}As was reported in earlier work \cite{JorRoesBro,Triques}, the exciton
COM~dispersions are highly non-parabolic, much like the hole subband dispersions.
This is not surprising, since the dispersion of the conduction subband is parabolic
and the hole mass is much larger than the electron mass. Furthermore, it has
been claimed in Ref.~\cite{Triques} that the exciton dispersion follows, in
a good approximation, the hole subband dispersion. Although this is certainly
true in the present case due to the parabolic electron dispersion and the small
electron to hole mass ratio, the exciton dispersion is in principle
a \emph{two}-particle quantity. In fact, the dispersion of the groundstate exciton
follows even more closely, in the studied cases within 1~meV, the electron-hole-pair
continuum edge \( {\cal E}_{n_{e}n_{h}}(\vec{Q}) \). The latter is defined by 
\begin{equation}
\label{XContEdge}
{\cal E}_{n_{e}n_{h}}(\vec{Q})=\underset {\vec{k}_{e}+\vec{k}_{h}=
 \vec{Q}}{min}\, \left\{ {\cal E}_{n_{e}}(\vec{k}_{e})+{\cal E}_{n_{h}}(\vec{k}_{h})\right\} \,,
\end{equation}
and represents
 the minimal kinetic energy of a free electron-hole pair for a given subband
combination \( n_{e}n_{h} \) and a given \( \vec{Q}. \) 
In the independent-subband approximation, this coincides with the 
\emph{exciton continuum edge}. 
Fig.~\ref{Fig_CONT_Contra_DISP}(a,b) directly compare for
a narrow and a wide QW the exciton groundstate dispersion and the appropriately
shifted exciton continuum edge. Also shown is the respective hole dispersion.
The latter lies always above the shifted exciton continuum edge since Eq.~(\ref{XContEdge})
implies \( {\cal E}_{n_{e}n_{h}}(\vec{Q})\leq {\cal E}_{n_{e}}(0)+{\cal E}_{n_{h}}(\vec{Q}) \).

The exciton groundstate dispersion is found to lie systematically 
below the shifted exciton
continuum edge, i.e., the groundstate exciton binding energy becomes larger
away from \( Q=0 \). This can be understood based on the fact that for \( Q\neq 0 \)
the groundstate exciton, Eqs.~(\ref{exciton1},\ref{exciton2}),
is built from hole
subband states around \( (1-\beta _{0})Q \). Due to the flatter subband dispersion
around this point, the hole mass gets larger and the wavefunction can
better adjust to the potential. The \( LH_{1}C_{1}-1s \) exciton in the 20~nm
wide QW, Fig.~\ref{Fig_CONT_Contra_DISP}b, is well separated from the spectrum
of the \( HH_{1}C_{1} \) exciton. 
We observe that at small \( Q \) the \( LH_{1}C_{1}-1s \) dispersion lies above 
the respective shifted exciton continuum edge. 
Indeed,
the \( LH_{1} \) subband shows a negative mass at the \( \Gamma  \)-point
which becomes positive near the avoided crossing. Hence, the respective exciton
has to pay with extra kinetic energy in order to achieve a small COM~momentum,
and its binding energy is decreased. For larger \( Q \) values it gains again
some binding energy.

\begin{figure}
\begin{center}
\resizebox*{8.5cm}{!}{\includegraphics{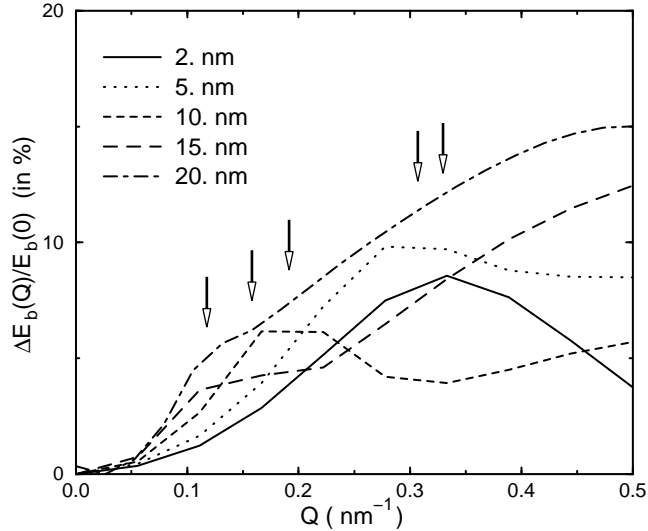}}
\end{center}

\caption{\label{Fig_Enhance_Bind}Enhancement of the exciton binding energy \protect\( E_{b}(Q)\protect \)
with increasing center of mass momentum \protect\( Q\protect \) for GaAs/Al\protect\( _{0.3}\protect \)Ga\protect\( _{0.7}\protect \)As
QW of various widths. The arrows mark the position of the respective \protect\( HH_{1}\protect \)-\protect\( LH_{1}\protect \)
avoided crossing, from the right to the left for growing QW width.}
\end{figure}
The enhancement of the groundstate binding energy for \( Q\neq 0 \) is particularly
large when the exciton is built from hole subband states around the avoided
crossings \( (1-\beta _{0})Q=k_{a.c.} \). This is demonstrated in Fig.~\ref{Fig_Enhance_Bind}
for QW of various widths. Peaks are seen at the respective location of the \( HH_{1} \)-\( LH_{1} \)
avoided crossing, marked by arrows. The enhancement of the groundstate binding
energy with \( Q \) is less than \( 15\% \) and is generally larger for wider
QW.

\begin{figure}
\begin{center}
\resizebox*{8.5cm}{!}{\includegraphics{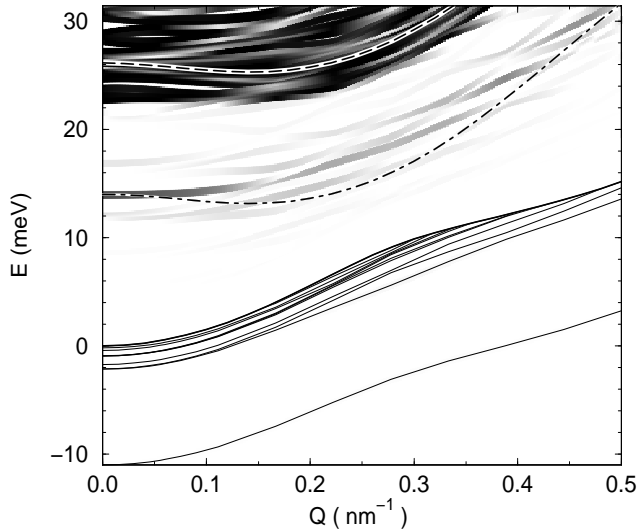}}
\end{center}

\caption{\label{Fig_sub_mix}Exciton dispersion for 5~nm GaAs/Al\protect\( _{0.3}\protect \)Ga\protect\( _{0.7}\protect \)As
QW. The \protect\( HH_{1}C_{1}\protect \) (thick) and \protect\( LH_{1}C_{1}\protect \)
(dashed) exciton continuum edges are also shown. The background shading indicates
how strong is the mixture of the \protect\( LH_{1}C_{1}\protect \) exciton
states to the numerically calculated eigenstates (black corresponds to 100\%).
For comparison, also a \protect\( LH_{1}C_{1}\protect \) (dot-dashed) exciton
continuum edge shifted to the energy of the \protect\( LH_{1}C_{1}-1s\protect \)
state at \protect\( Q=0\protect \) has been drawn.}
\end{figure}
In Fig.~\ref{Fig_sub_mix}, we plot the exciton dispersion of a 5~nm QW. The
exciton dispersions, like the hole subband dispersions, are less anharmonic
in the much narrower QW than for the QW of Fig.~\ref{Fig_k-disp} because of
the larger energy separation of the hole subbands. The avoided crossing takes
place at rather large \( k \) (\( k=0.31 \)~nm\( ^{-1} \)). The shading indicates
the percentage of the contribution of the \( LH_{1}C_{1} \) subband states
to the norm of the numerical eigenvectors. The very light shading of the bound
exciton states confirms the small admixture of the \( LH_{1}C_{1} \) states
in the groundstate. The shading is darker in the vicinity of the avoided crossing.
Small-scale intensity variations are numerical artefacts due to the finite \( k \)-space
mesh. In the exciton continuum only the shading is shown. The \( LH_{1}C_{1}-1s \)
resonance is clearly seen starting at approximately 18~meV. The resonance is
sharper at the beginning and becomes more diffuse at the avoided crossing. It
lies slightly above the respective exciton continuum edge.

\subsection{Real-space calculations\label{RSECTION}}

For effectively two-dimensional structures with translational symmetry like
the considered symmetric QW in axial approximation and, maybe, for some highly
idealized quantum wire structures, the real-space method presented in section
\ref{REALSPACE} can not compete with the one in \( k \)-space. 
However, for realistic
one-dimensional structures, like V-groove and T-shaped quantum wires, this may
be the only feasible approach for calculating the exciton groundstate dispersion.
This is due to the high number of confining dimensions for the exciton (in quantum
wires four, two for each particle): an expansion in a problem-adapted basis
like the product basis of the one-particle eigenstates, for which one expects
reasonable convergence, leads to four-dimensional integrals for the Coulomb
interaction. An expansion in a basis where the Coulomb potential is simple 
will probably show a very slow convergence with basis size. 

The calculations reported here are mainly to be viewed as tests of the applicability
of our real-space approach and its generalization to finite-elements discretization.
They are primarily compared with results obtained with the more established
\( k \)-space methods. We will therefore discuss the results of the real-space
calculations focusing on the convergence properties of the method.
Further, wavefunction features are better visualized in real space, 
in particular, the electron-hole correlation in the growth direction. 

For the real-space calculations at finite \( Q \), we used the optimized \( \beta _{0} \)
obtained in the respective \( k \)-space calculations. We could have used equally
well some other procedure to find the effective Bohr radius \( a_{B} \), e.g.,
a variational one, or we could also have fitted a 2D \( 1s \)-exciton function
to the in-plane probability distribution of the previously calculated exciton
groundstate at \( Q=0 \). 

The Coulomb potential was discretized as described in the Appendix. The 3D \(1s\)-exciton
in the four dimensional space (\( \vec{\rho },z_{e},z_{h} \))
 with \( m^{ref}_{e} =  0.0665\, m_{0} \),
\( m^{ref}_{h} = 0.24\, m_{0} \) was used as reference groundstate.
The value for the reference hole mass was taken from Fig.~\ref{Fig_Massen},
discussed below, as an average value for the range of QW widths considered.
This gives a reasonable reference Bohr radius of \( a^{ref}_{B}=12.2 \)~nm;
it is nearly the correct value for the in-plane motion or somewhat larger. In
the confinement direction the size of the reference wavefunction is larger than
the actual one (the exciton is quenched in this direction), too. As discussed
in the Appendix, a reference Bohr radius as large as or somewhat larger than
the actual one gives good convergence. We did test calculations with \( a_{B}^{ref} \)
doubled and with \( a_{B}^{ref} \) halved and found a qualitatively similar
behavior as in Fig.~\ref{Fig_Glutsch} in the Appendix. The integration region
was 120~nm wide in the \( \vec{\rho } \) directions and 30~nm (60~nm) wide in
the \( z \) directions for the 5~nm (20~nm) QW with a gridpoints distance of 
about \( a_{B}/6 \) in the in-plane directions.

\begin{figure}
\begin{center}
\resizebox*{8.5cm}{!}{\rotatebox{-90}{\includegraphics{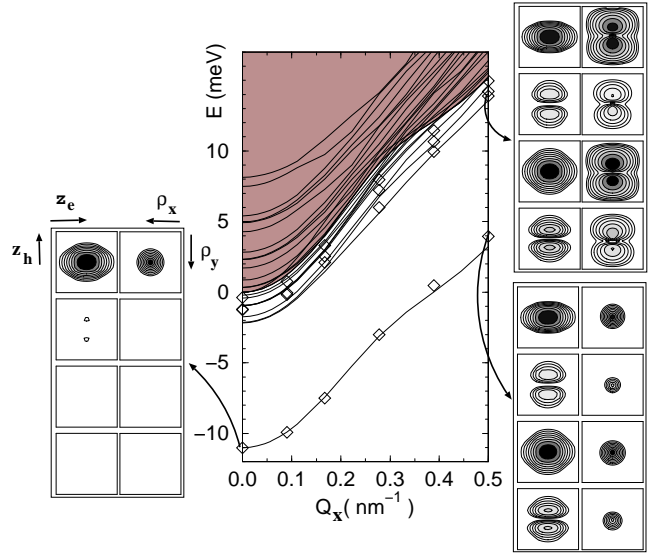}}}
\end{center}

\caption{\label{Fig_rho_plots}Middle panel: Exciton dispersion for a 5~nm wide QW.
$k$-space results for the groundstate and some excited 
bound and continuum states (lines) and the exciton continuum 
(shaded) are presented, as well as some real-space results (diamonds). 
Side panels with
logarithmic contour plots of the exciton probability distribution for each spin-component
(from top to bottom \protect\( m_{J}=\textstyle +\frac{3}{2},+\frac{1}{2},-\frac{1}{2},-\frac{3}{2}\protect \))
are shown for characteristic exciton states 
(thick lines) at \protect\( Q=0\protect \) (left)
and \protect\( Q_{x}=0.5\protect \)~nm\protect\( ^{-1}\protect \) 
(right panels). The
displayed area is in each direction \protect\( 10\protect \)~nm wide for the
\protect\( z_{e}z_{h}\protect \) and about \protect\( 10\, a_{B}\protect \)
wide for the \protect\( \rho _{x}\rho _{y}\protect \) plots, respectively.}
\end{figure}
The panel in the middle of Fig.~\ref{Fig_rho_plots} shows dispersions of the
lowest exciton states (diamonds) from the real-space calculation for a 5~nm wide
QW. The exciton dispersions calculated in momentum space are plotted as full
curves for comparison. The groundstate binding energy is not yet fully converged:
all real-space results were shifted approximately 1~meV to lower energies to
match the groundstate energies at \( Q=0 \) for both methods. Numerical tests
show that the gridpoint density in the growth direction ($\approx 0.6$ points/nm)
is more critical than in the in-plane direction ($\approx 0.6$ points/nm).
The exciton
continuum edge lies also 0.5~meV too high, and the stronger confinement due to
the Coulomb interaction lets us expect for the exciton a larger deviation. Nevertheless,
the groundstate relative dispersion, \( {\cal E}_{g}^{X}(Q)-{\cal E}_{g}^{X}(0) \),
is converged and reproduces the \( k \)-space results very well. We note
that in the case of parabolic one-particle dispersions, this is an exact property
of any numerical exciton dispersions. The dispersions of the excited states 
are not reproduced that well. This is mainly due to their larger spatial extension
and smaller energy separation from each other compared to the groundstate.

The panels on the left and on the right in Fig.~\ref{Fig_rho_plots} show logarithmic
contour plots of the exciton probability distribution for some characteristic
states. The probability distribution is either integrated over \( z_{e},\, z_{h} \)
and displayed in the \( \vec{\rho } \)~plane, or integrated over~\( \vec{\rho } \)
and displayed in the \( z_{e}z_{h} \)~plane for each spin component separately.
The numerically obtained wavefunctions are a linear combination of the two degenerate
solutions with opposite parity, Eq.~(\ref{-P2}). 
They are disentangled according to parity \( P \), and only \( P=-1 \) states are displayed.

The left panel displaying the groundstate exciton at \( Q=0 \) illustrates
its \( HH_{1}C_{1}-1s \) character: the main spin component is \( m_{J}=+3/2 \)
and has no nodes. The bulk of the exciton is confined in the QW but there is
substantial penetration into the barrier, being stronger for the lighter electron
(note the logarithmic plot). At \( Q=0.5 \)~nm\( ^{-1} \) (right panels),
the groundstate exciton has still \( HH_{1}C_{1}-1s \) character, as is seen
in the \( k \)-space calculations, Fig.~\ref{Fig_k-envelopes}. This does not
contradict the strong mixture of heavy- and light-hole bulk states seen in the
lower right panel of Fig.~\ref{Fig_rho_plots}. 
Indeed, the exciton is built from hole subband states near
\( (1-\beta _{0})Q \). This point lies past the \( HH_{1}-LH_{1} \) avoided
crossing. Hence, the \( HH_{1} \) subband states near this point are a strong
mixture of heavy- and light-hole bulk bandedge states. The stronger penetration
into the barrier of the light-hole component is again related to its smaller
mass. 

An interesting feature is the larger confinement of the \( m_{J}=\textstyle +\frac{3}{2}\displaystyle  \)
component at \( Q=0.5 \)~nm\( ^{-1} \) compared to \( Q=0 \). This is a consequence
of the enhanced exciton binding energy. Altogether, the plots demonstrate that
the total charge distribution in not altered much with increasing COM~momentum;
the somewhat stronger confinement of the heavy hole is at least partly canceled
by the larger penetration into the barrier of the light hole. Although here
not clearly resolved, the in-plane plots at \( Q=0.5 \)~nm\( ^{-1} \) show the
slight deformation of the originally radially symmetric wavefunction that was
seen in Fig.~\ref{Fig_SHIFT_k}. Again these deformations partly cancel each
other in the sum over the spin components, and the in-plane charge distribution
remains mainly symmetric. This is more clearly seen for the first excited state
at \( Q=0.5 \)~nm\( ^{-1} \) which has \( HH_{1}C_{1} \)-\( 2p_{y} \) character
(upper right panel in Fig.~\ref{Fig_rho_plots}). It is the \( HH_{1}C_{1}-2p_{+} \)
exciton, which is, again, slightly lower in energy than the \( HH_{1}C_{1}-2p_{-} \)
exciton at \( Q=0 \). At large \( Q \) the character of the \( HH_{1}C_{1} \)
envelope changes from \( 2p_{\pm } \) to \( 2p_{y,x} \).

\begin{figure}
\begin{center}
\resizebox*{8.5cm}{!}{\includegraphics{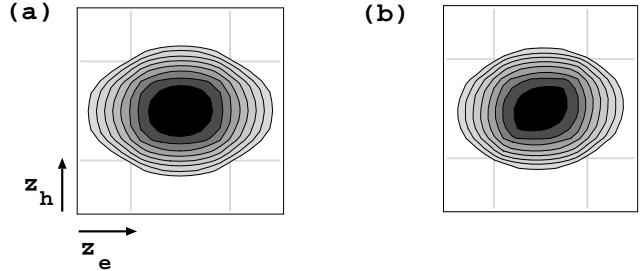}}
\end{center}

\caption{\label{Fig_ze_zh_corr_magn}Groundstate exciton probability density in the
\protect\( z_{e}z_{h}\protect \)-plane for the main spin-component (\protect\( m_{J}=\textstyle +\frac{3}{2}\protect \))
at \protect\( Q=0\protect \) for a 5~nm wide QW, (a) integrated over the in-plane
relative coordinate \protect\( \vec{\rho }\protect \) (same as lower left corner
of Fig.~\ref{Fig_rho_plots}) and (b) at \protect\( \vec{\rho }=0\protect \).
The background lines mark the position of the QW interfaces.}
\end{figure}
One does not expect strong electron-hole correlation in the growth direction
for the 5~nm QW, which is considerably narrower than the exciton Bohr radius.
Here, 
the confining potentials are on average much stronger than the Coulomb potential
and the wavefunction can not relax in this direction. 
Indeed, in Fig.~\ref{Fig_ze_zh_corr_magn}(a)
the probability density integrated over \( \vec{\rho } \) does not show much
correlation: the contour lines are not elongated 
along the diagonal, \( z_{e}=z_{h} \). 
However, some correlation exists as the plot of the cut at \( \rho =0 \) in
Fig.~\ref{Fig_ze_zh_corr_magn}(b) demonstrates. This weak correlation is not
included in our \( k \)-space calculations where only one conduction band was
used. Its envelope does not depend on \( k \) and consequently the \( z_{e} \)
coordinate can be separated, Eq.~(\ref{exciton1}). 

\begin{figure}
\begin{center}
\resizebox*{8.5cm}{!}{\rotatebox{-90}{\includegraphics{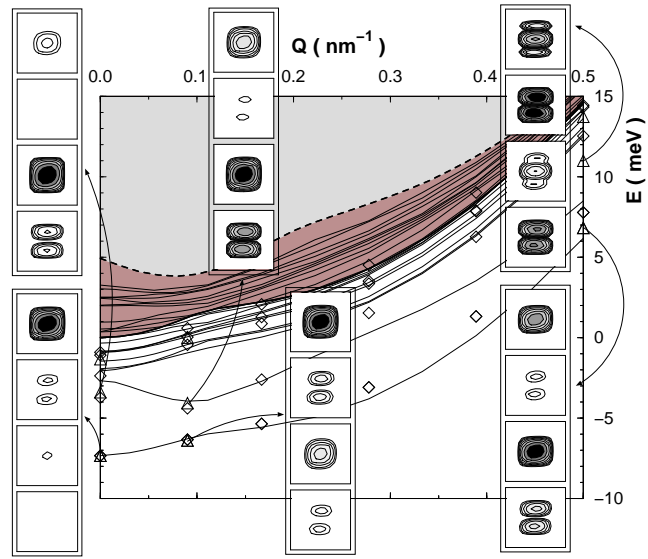}}}
\end{center}

\caption{\label{Fig_ze_zh_corr}Logarithmic contour plots of the exciton probability
distribution in the \protect\( z_{e}z_{h}\protect \)-plane for each spin-component
(in each panel from top to bottom \protect\( m_{J}=\textstyle +\frac{3}{2},+\frac{1}{2},-\frac{1}{2},-\frac{3}{2}\protect \))
for the \protect\( HH_{1}C_{1}-1s\protect \) and the \protect\( LH_{1}C_{1}-1s\protect \)
excitons at some characteristic \protect\( Q\protect \)-values for a 20~nm QW.
The displayed region is in each direction twice the QW width. The triangles
(diamonds) show results of the real-space calculations for a dense (less dense)
mesh. 
Lines and shading of the background panel are the \protect\( k \protect \)-space
results of Fig.~\ref{Fig_k-disp}.
We do not show the \protect\( HH_{2}C_{1}\protect \) continuum that lies partly
in the displayed region.}
\end{figure}
In Fig.~\ref{Fig_ze_zh_corr}, results are displayed for a 20~nm wide QW. Two
set of points are shown for two different mesh sizes. The energies of the groundstate
exciton are almost converged for the more dense mesh; the deviation from the
results of the \( k \)-space calculations is only 0.2~meV at \( Q=0 \). 

The small, but noticeable \( m_{J}=+1/2 \) component of the groundstate exciton
\( HH_{1}C_{1}-1s \) exciton in the lower left panel corresponds to the substantial
admixing of the \( LH_{1}C_{1}-d_{-} \) exciton seen already for the 15~nm QW
in Fig.~\ref{Fig_k-envelopes}. The admixing is larger for wider QW due to the
smaller energy separation of the respective subbands. At \( Q=0.5 \)~nm\( ^{-1} \),
it has mainly bulk light-hole character (\( m_{J}=-1/2 \)).
 This, again, does
not contradict the \( HH_{1}C_{1}-1s \) character since the \( HH_{1} \) subband
has approximately 60\% light-hole character
beyond the avoided crossing with the \( LH_{1} \) subband. 
The first excited state at \( Q=0.5 \)~nm\( ^{-1} \)
 (\( LH_{1}C_{1}-1s \) exciton) has \( LH_{1}C_{1}-1s \) character
even though an additional node is seen in the \( z_{e}z_{h} \) probability
distribution of the main spin component (\( m_{J}=+1/2 \)). 
The envelopes of
the single-particle subband states at large enough in-plane momentum show more
nodes than at the \( \Gamma  \)-point due to the coupling of in-plane and growth
directions in the Luttinger Hamiltonian. This is one of the reasons why the
expansion Eqs.~(\ref{exciton1},\ref{exciton2}) gives very good results with
just two subbands, while an expansions in the subband states at the \( \Gamma  \)-point
\cite{Triques} needs more subbands for the same accuracy.

The probability distribution plots illustrate the almost vanishing penetration
into the barriers for the wide QW, in contrast to the narrower QW of Fig.~\ref{Fig_rho_plots}.
All panels show a clear orientation of the contour lines towards the
diagonal,  \( z_{e}=z_{h} \).
This demonstrates the considerable electron-hole correlation
in the growth direction for QW wider than one Bohr radius. Still, the stronger
correlation has little impact on the energies. Recall that in perturbation theory
the first order correction to the wavefunction gives only a second order correction
to the energy. This justifies the usual factoring out of the dependence on the
growth direction for the much lighter electron.

This, however, is not the case for the much heavier hole. It is a well known
fact, that the Coulomb coupling of the hole subbands is considerable. Indeed,
for the 20~nm wide QW, neglecting the Coulomb coupling of the \( HH_{1}C_{1} \)-
to the \( LH_{1}C_{1} \)-excitons leads to an error in the groundstate binding
energy larger than \( 10\% \). That is, the correlation of in-plane and confinement
directions for the hole is substantial.

For the 20~nm QW, a single calculation at \( Q=0.5 \)~nm\( ^{-1} \) was done
with \( \beta =1 \) in order to check the relevance of this parameter for the
numerical accuracy in the real-space calculations, too. Indeed, the respective
groundstate energy lies very far (\,43~meV\,!\,) above the correct value.

\subsection{Average exciton groundstate mass\label{MASSSECTION}}

In the process of determining the optimal choice of the COM~coordinate system,
we derived in section \ref{CMM} an expression for the kinetic mass of the groundstate
exciton, Eqs.~(\ref{newMX},\ref{effmassSA}). Two assumptions were essential:
First, the Coulomb potential, i.e., the form factors, are a function of the in-plane
momentum transfer, and, second, the Ansatz (\ref{shiftA}) is valid. The numerical
results of the previous sections support these assumptions.

Together with the exciton dispersions in Fig.~\ref{Fig_k-disp}, we displayed
for the groundstate exciton a parabola with the exciton mass \( M_{g}^{X} \)
of section \ref{CMM}. This mass is obviously not the one determined by the curvature
of the exciton groundstate dispersions at \( Q=0 \). It is rather an average
of the curvature of the groundstate dispersion in a region of size 
\( 1/a_{B} \). Indeed, this is implied by Eqs.~(\ref{newMX},\ref{effmassSA})
and the observation made in section \ref{KSECTION} that the groundstate dispersion
follows closely the respective exciton continuum edge. 

This exciton COM~momentum region is the one important for exciton localization
due to well width fluctuations, interface roughness or alloy fluctuations
in QW \cite{Runge}. Indeed, the exciton averages over smaller scale fluctuations
due to its finite size and feels an effective disorder potential that is spatially
correlated over the Bohr radius \( a_{B} \). 

\begin{figure}
\begin{center}
\resizebox*{6.5cm}{!}{\includegraphics{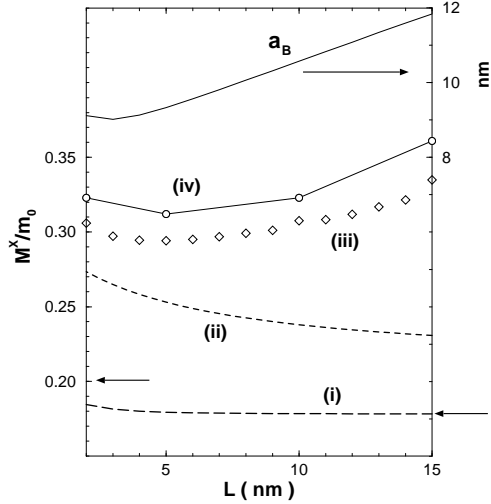}}
\end{center}

\caption{\label{Fig_Massen}Exciton average kinetic mass in various approximations and
Bohr radius \protect\( a_{B}\protect \) versus well width \protect\( L\protect \)
for GaAs/Al\protect\( _{0.3}\protect \)Ga\protect\( _{0.7}\protect \)As QW.
The figure is discussed in section \ref{MASSSECTION}. }
\end{figure}
The dependence of various expressions for the groundstate exciton mass on the
QW width \( L \) is displayed in Fig.~\ref{Fig_Massen}. The conduction band
mass was taken material-independent \( m^{b,\, w}_{e}=0.0665 \), the shown \( L \)-dependence
comes solely from the valence band. Displayed are the masses obtained by: (i)
describing the hole in the diagonal Luttinger approximation \( 1/m_{h}=P_{w}\, (\gamma ^{w}_{1}+\gamma ^{w}_{2})+P_{b}\, (\gamma _{1}^{b}+\gamma _{2}^{b}) \),
where \( P_{w,b} \) denote the probability that the hole is in the well and
barrier material respectively (long dashed), (ii) taking as hole mass the subband
curvature at the \( \Gamma  \)-point that is known analytically \cite{Foreman}
(dashed), (iii) using our semi-analytical expression 
 (\ref{newMX},\ref{effmassSA}) for the mass
with a 2D \(1s\)-exciton function fitted to the envelope of the \( HH_{1}C_{1} \)
component (diamonds), and (iv) the average of the curvature of the numerical
exciton dispersion weighted with the same function as (iii) (circles, the line
is a guide to the eye). 

The top curve displays the values for the Bohr radius that we have used for the calculation
of (iii) and (iv). Arrows at either sides of the lower part of Fig.~\ref{Fig_Massen}
mark the mass in the diagonal approximation in the well and barrier bulk materials. 

Fig.~\ref{Fig_Massen} demonstrates the failure of the diagonal Luttinger approximation
(i) to describe even the \( HH_{1} \) subband curvatures at the \( \Gamma  \)-point
due to the degeneracy of heavy- and light-hole bands in the unstrained bulk.
However, even the correct single-particle subband curvatures at the \( \Gamma  \)-point
(ii) fail to describe accurately the curvature of the groundstate exciton dispersion
at \( Q=0 \). This is mainly due to the finite extension of the exciton in
\( k \)-space that implies an averaging of the subband dispersions over a region
of approximately \( 1/a_{B} \) near the \( \Gamma  \)-point and partly to
the Coulomb coupling to higher subbands. Both effects tend to make the groundstate
exciton heavier. The mass derived from the curvature of the groundstate exciton
at \( Q=0 \) (not shown) lies between curves (ii) and (iii).

The numerically obtained ``best'' mass values (iv) show a quantitatively
and qualitatively different behavior from curves (i) and (ii). For very narrow
QW the subbands become flatter at the \( \Gamma  \)-point because of the larger
penetration into the barriers where the exciton becomes again heavier, as in
models (i), (ii). But, for large \( L \) the region of the \( HH_{1} \)-\( LH_{1} \)
avoided crossing comes to a distance of approximately \( 1/a_{B} \) to the
\( \Gamma  \)-point and the exciton, averaging over the flatter subband dispersion,
becomes heavier. 

The quality of our semi-analytical result for the average exciton mass (iii)
has to be judged according to its deviation from the numerical average (iv).
The non-monotonous behavior is clearly seen for the mass (iii) obtained using
only the \( HH_{1} \) subband dispersion and the fitted Bohr radius. The mass
values (iii) are somewhat smaller (\(<\)10\%) than the ones of curve (iv). This
is due to the enhancement of the binding energy for \( Q\neq 0 \) that yields
larger average masses (iv) than one would expect based on the one-particle subband
dispersions and the \( Q=0 \) groundstate exciton. Our semi-analytical average
groundstate exciton mass (iii) reaches a minimum approximately at the QW width
where the maximum binding energy is reached. 

The small differences between curves (iii) and (iv) demonstrate the quality
of our expression (\ref{newMX},\ref{effmassSA}) for the average exciton
groundstate mass. For this reasonable and easy-to-use mass expression, only a
good estimate for the in-plane Bohr radius and the dispersion of the involved
single-particle subbands is needed. 

In the paper by Triques and Brum \cite{Triques}, average exciton effective
masses were calculated that are relevant to the formation process of excitons
in two different scenarios. These masses are defined by parabolic fits within
a relevant energy range of: (a) 5~meV, approximately half the exciton binding
energy, in the case that the particles first relax and then form an exciton
of kinetic energy lower than the binding energy and (b) 36~meV in the case that
the exciton is formed very fast and relaxes initially via optical phonon emission,
reaching energies below 36~meV, the energy of the GaAs-LO phonon. These energy
ranges translate to \( Q \) values in general much larger than \( 1/a_{B} \).
Therefore, these average masses should be larger than those of the present work.
However, the values published in \cite{Triques} for the 5~meV mass (\( 0.2m_{0}\leq M_{g}^{X}\leq 0.3m_{0} \)
for \( L\leq 10 \)~nm) are smaller than ours for narrow QW. Our average masses
of scenario (a) are not falling below 0.3\( m_{0} \) and show a smooth minimum
for a QW of width somewhere between 2~nm and 5~nm. This difference can be traced
back mainly to the inefficiency for narrow QW of their method involving an expansion
in the subband states at the \( \Gamma  \)-point, as already remarked by the
authors themselves. Indeed, for narrow QW the few states at the \( \Gamma  \)-point
can not provide the needed flexibility to simulate states far away from the
\( \Gamma  \)-point. This is substantiated by the lower left panel in Fig.~\ref{Fig_rho_plots}
showing the groundstate exciton at \( Q=0.5 \)~nm\( ^{-1} \)for the 5~nm QW.
The \( z_{e}z_{h} \)-plots show a substantial \( m_{J}=+\textstyle \frac{1}{2} \)
spin component. In an expansion of the \( z_{h} \) dependence in the hole subband
states at the \( \Gamma  \)-point this component would need a \( LH_{2} \)
subband to be described efficiently. However, for the 5~nm QW only the \( HH_{1},\, LH_{1},\, HH_{2} \)
subbands exist below the top of the barrier. 

In Fig.~\ref{Fig_CONT_Contra_DISP}(c), we compare the results of Ref.~\cite{Triques}
with ours for the 5~nm QW. For this narrow QW, their exciton and \( HH_{1} \)
subband dispersions are much steeper than ours. The discrepancy is not due to
the different parameters as a comparison with the subband dispersion calculated
exactly with the transfer matrix method for the parameters of Ref.~\cite{Triques}
demonstrates (diamonds). 

We remarked already, that different parts of the non-parabolic groundstate
exciton dispersion are relevant for different physical process, and 
we gave exciton localization \cite{Runge,Hess} and cooling of a non-thermal exciton 
population  \cite{Triques} as examples.
We would like to discuss two more possible experimental consequences
of anharmonicities using Fig.~\ref{Fig_k-disp} as illustration. 
First, an exciton population loosing energy by acoustic phonon emission at
low temperatures could experience a bottle-neck effect, i.e., an increased
population of the $k$-space region  near $0.15$~nm$^{-1}$. 
Scattering closer towards the $\Gamma$-point will be suppressed by the decreased
density of final states. 
A second, more directly observable consequence is the temperature dependence 
of the exciton lifetime. Evaluation of the latter along the lines of 
Ref.~\cite{And95} with the groundstate dispersion of Fig.~\ref{Fig_k-disp}
yields a super-linear increase of the exciton lifetime
with temperature: The slope increases in the range from 5 to 20~K 
by a factor~2.2 (not shown).

\section{Concluding remarks\label{Sconclude}}

In summary, we performed \( \vec{k}\cdot \vec{p} \) multiband exciton dispersion
calculations of high accuracy even for very large COM~momentum and narrow
QW using the well known expansion in subband states. For the high quality of
the numerical results, an optimized COM~transformation was essential. The
optimized COM~transformation is related to an average exciton kinetic mass
for which a simple semi-analytical expression is derived. This eliminates for
many practical purposes the need to actually calculate the groundstate dispersion.
The groundstate exciton dispersion is found to follow closely the respective
exciton continuum edge.

In addition, we demonstrated that multiband groundstate exciton dispersion calculations
are feasible in real space using a finite-differences scheme. Besides the essential
optimized COM~transformation, a convenient groundstate-adapted discretization
of the Coulomb potential enhances numerical accuracy. This method promises to
give results for systems where the first one is not practicable. Its generalization
for finite-element schemes is straightforward.

\section*{Acknowledgments}

This work has been funded by the Deutsche Forschungsgemeinschaft in the frame
of SFB 296. We wish also to thank the Rechenzentrum of the Humboldt University
and the Konrad-Zuse-Zentrum in Berlin for their support and access to their
Cray J932 (project bvph08as).

\section*{Appendix: Discretization of the Coulomb potential\label{VCdiscr}}

In order to attain a simple discretization for the interaction of some Hamilton
operator, Glutsch, Chemla, and Bechstedt \cite{Glutsch} proposed to discretize
on the same mesh another operator whose groundstate is analytically known with
the same interaction but with a simple kinetic term (mass \( m^{ref} \)). 
We will call this
the reference system. For illustration, consider a simple one-dimensional system
with known groundstate wavefunction \( g(x) \) of energy \( E^{ref}_{g} \).
The Schr\"odinger equation of the reference system 
discretized on
a mesh  \( x_i = i\cdot \Delta _{x} \) reads 
\begin{equation}
\label{VCdiscr1}
-\frac{\hbar ^{2}}{2m^{ref}\, \Delta _{x}^{2}}\left( g(x_{i+1})-2g(x_{i})+g(x_{i-1})\right) +V(x_{i})\, g(x_{i})=E^{ref}_{g}\, g(x_{i})
\end{equation}
and  yields the groundstate adapted discretization 
\begin{equation}
\label{GlutschVC}
V(x_{i})=E^{ref}_{g}+\frac{\hbar ^{2}}{2m^{ref}\,
 \Delta _{x}^{2}}\frac{g(x_{i+1})-2g(x_{i})+g(x_{i-1})}{g(x_{i})}\, ,\, \, g(x_{i})\neq 0\,\,.
\end{equation}
This procedure is very simple, easy to implement, cheap to calculate, and gives
for the reference system always the correct groundstate energy, regardless how
inappropriate the mesh is. In addition, no special handling for potentials with
integrable singularities is needed.
\begin{figure}
\begin{center}
\resizebox*{8.5cm}{!}{\rotatebox{-90}{\includegraphics{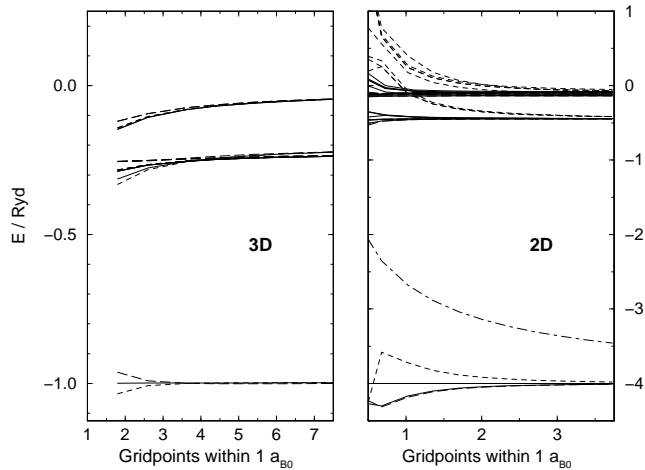}}}
\end{center}

\caption{\label{Fig_Glutsch}The lowest eigenvalues calculated in real space of the
ideal 3D exciton (left) and 2D exciton (right) of Bohr radius \protect\( a_{B0}\protect \)
plotted versus grid density. Discretization of the Coulomb potential as described
in the Appendix with reference Bohr radius \protect\( a^{ref}_{B}\protect \):
\protect\( a_{B0}\protect \)(thick), \protect\( 2\, a_{B0}\protect \)(dashed),
\protect\( a_{B0}/2\protect \) (long dashed). 
For the 2D case, also results are shown for the groundstate   
with the potential integrated analytically in every mesh cell, Eq.~(\ref{Coul2DInt}),
(thin) and for the groundstate with the ``naive discretization'' (dot-dashed).
The former curve is hardly to distinguish from the one for \protect\( a_{B}^{ref}=2\, a_{B0}\protect \).}
\end{figure}

Using this discretization for a reference state similar enough to the one sought,
one can expect good convergence with mesh size. In order to check how the dependence
of the potential discretization on the reference state influences the results,
we performed calculations of the ideal 2D and 3D excitons using discretizations
of the Coulomb potential based on reference excitonic groundstates with various
Bohr radii. Fig.~\ref{Fig_Glutsch} shows the lowest numerical eigenvalues as
a function of gridpoint density. One can see that the correct estimation of
the ``unknown'' groundstate is not so critical: a reference Bohr radius \( a_{B} \)
within a factor of two from the actual one, \( a_{B0} \), still gives good
results for the groundstate for reasonable mesh densities. Apparently, especially
for the 2D case, it is better to choose \( a_{B} \) rather larger than smaller
in order to get good results also for the excited states. 

We also show for the 2D case the results obtained with the potential integrated
analytically on each Cartesian mesh element, using 
\begin{equation}
\label{Coul2DInt}
\int \! \! \! \int \frac{dx\,\,dy}{r}=
y \, \ln \left( x+\sqrt{x^{2}+y^{2}}\,\right) +
x \, \ln \left( y+\sqrt{x^{2}+y^{2}}\,\right)\,. 
\end{equation}
 This  gives also a very good convergence with the mesh size. The 2D
result obtained for the groundstate with the potential integrated analytically
only at the origin, and taking everywhere else its value at each gridpoint ('naive'
discretization) is included in Fig.~\ref{Fig_Glutsch}. Since both discretizations
are the same at the origin the difference does not originate from the divergence
at this point. The superiority of the discretizations (\ref{GlutschVC}) and
(\ref{Coul2DInt}) is obvious. We therefore expect Eq.~(\ref{GlutschVC}) to yield
good results even with not very dense meshes.

For the 3D case a similar expression to Eq.~(\ref{Coul2DInt}) for the Coulomb
potential integrated analytically on a rectangular box can be derived. However,
we could not use this result to attain an alternative discretization for the
real-space QW calculations because the natural mesh \( \left( \rho _{x},\rho _{y},z_{e},z_{h}\right)  \)
is not Cartesian in the relative coordinate \( z_{e}-z_{h} \).

%%%%%%%%%%%%%%%%%%%%%%%%%%%%
%\end{multicols}
%%%%%%%%%%%%%%%%%%%%%%%%%%


\begin{thebibliography}{}
\bibitem{US2}Unfortunately, in Fig.~1 of Ref.~\cite{US} a different convention was used,
with \( \beta  \) replaced by \( \beta -1/2 \).
\end{thebibliography}
\end{document}